# A Study of Transverse Momentum Distributions of Hadrons at LHC

**Saeed Uddin [*], Inam-ul Bashir and Riyaz Ahmed Bhat**

*Department of Physics, Jamia Millia Islamia (Central University)
New Delhi-110025*

## Abstract

The transverse momentum distributions of various hadrons produced in most central Pb+Pb collisions at LHC energy $\sqrt{s_{NN}}$ = 2.76 TeV have been studied using our earlier proposed unified statistical thermal freeze-out model. The calculated results are found to be in good agreement with the experimental data measured by the ALICE experiment. The model calculation fits provide the thermal freeze-out conditions in terms of the temperature and collective flow effect parameters for different particle species. Interestingly the model parameter fits reveal a strong collective flow in the system which appears to be a consequence of the increasing particle density at LHC. The model used incorporates a longitudinal as well as transverse hydrodynamic flow. The chemical potential has been assumed to be nearly equal to zero for the bulk of the matter owing to a high degree of nuclear transparency effect at such energies. The contributions from heavier decay resonances are also taken into account in our calculations.

## 1. Introduction

The study of identified particle spectra in heavy-ion collisions at ultrarelativistic energies helps us to learn about the final state distribution of baryon numbers among various particle species at the thermo-chemical freeze-out after the collision which is initially carried by the nucleons only [1].

Within the framework of the statistical model it is assumed that a hot and dense fireball is formed over an extended region for a brief period of time after the initial collision and it undergoes collective expansion leading to a decrease in its temperature and finally to the hadronization. After the hot fireball formed in such collisions hadronizes, the hadrons keep rescattering with each other and continue to build up collective expansion. Consequently, the matter becomes dilute and the average distance between hadrons exceeds the range of the strong interactions. At this point of time, all scattering processes stop and the hadrons freeze out [2].

The chemical freeze-out occurs earlier when the rates for inelastic processes, in which secondary hadrons are produced or the hadrons change their identity become too small to keep up with the expansion. Since the corresponding inelastic cross sections are only a small fraction of the total cross section at lower energies, the inelastic processes stop well before the elastic ones, leading to an earlier chemical freeze-out for the hadron abundances. Finally at a later stage the hadrons completely decouple from each other such that even the elastic processes also come to a halt. Consequently, the momentum spectra get frozen in time and a thermal freeze-out occurs. Thus chemical freeze-out precedes thermal or kinetic freeze-out [3].

The identified particle spectra provide information, both about the temperature of the system and the collective flow at the time of thermal freeze-out. Collective flow depends on the internal pressure gradients created in the collision and is addressed by hydrodynamic models [4, 6]. These effects are species-dependent. The produced hadrons are believed to carry information about the collision dynamics and the subsequent space-time evolution of the system.

Hence an accurate measurement of the transverse momentum distributions of identified hadrons along with the rapidity spectra is essential for the understanding of the dynamics and the properties of the created matter up to the final thermal or hydrodynamical freeze-out in case of collective flow [7].

It has been shown earlier [2] that this model successfully explains the rapidity and transverse momentum distributions of hadrons and their ratios in Au+Au collisions at highest RHIC energy of $\sqrt{s_{NN}}$ = 200 GeV. In this paper, we briefly describe the model and use it to reproduce the transverse momentum distributions of hadrons produced in Pb+Pb collisions at $\sqrt{s_{NN}}$ = 2.76 TeV.

## 2. Model

In order to obtain the particle spectra in the *overall rest frame of the hadronic fireball* in our model we first define the invariant cross-section for given hadronic specie in the local *rest frame of a hadronic fluid element*. Since the invariant cross section will have the same value in all Lorentz frames [8] we can thus write,

$$E \frac{d^3N}{d^3p} = E' \frac{d^3N}{d^3p'} \qquad (1)$$

Where $E$ ($E'$) is the energy of the particle and $p$ ($p'$) is the momentum. The primed quantities on the RHS refer to the invariant spectra of given hadronic specie in the rest frame of the local hadronic fluid element, while the unprimed quantities on the LHS refer to the invariant spectra of the same hadronic specie in the overall rest frame of the hadronic fireball. The occupation number distribution of the hadrons in the momentum space follows the distribution function

$$E' \frac{d^3N}{d^3p'} \sim \frac{E'}{e^{(\frac{E'-\mu}{T})} \pm 1} \qquad (2)$$

where (+) sign and (-) sign are for fermions and bosons, respectively, and $\mu$ is the chemical potential of the given hadronic specie. For the temperatures under consideration and the large masses of hadrons it is safe to work with Boltzman distribution.

In recent works [7, 9] it has been clearly shown that there is a strong evidence of increasing baryon chemical potential, $\mu_B$ along the collision axis in the RHIC experiments. In view of this fact we write the expression for chemical potential as $\mu_B =$ $a + b\, y_0^2$ [7, 9, 10], where $y_0$ is the rapidity of the expanding hadronic fluid element. Further it is assumed [7] that the rapidity of the expanding hadronic fluid element $y_0 \propto z$ or $y_0 = \xi\, z$, where $z$ is the longitudinal coordinate of the hadronic fluid element and $\xi$ is a proportionality constant. The above conditions also ensure that under the transformation $z \to -z$, we will have $y_0 \to -y_0$, thereby preserving the symmetry of the hadronic fluid flow about $z = 0$ along the rapidity axis in the centre of mass frame of the colliding nuclei.

The *transverse* velocity component of the hadronic fireball, $\beta_T$ is assumed to vary with the transverse coordinate $r$ in accordance with the Blast Wave model as $\beta_T(r) = \beta_T^s \left(\frac{r}{R}\right)^n$ [11], where $n$ is an index which fixes the profile of $\beta_T(r)$ in the transverse direction and $\beta_T^s$ is the hadronic fluid *surface transverse expansion velocity* and is fixed in the model by using the parameterization $\beta_T^s = \beta_T^0 \sqrt{1 - \beta_z^2}$ [7]. This relation is also required to ensure that the net velocity $\beta$ of *any* fluid element must satisfy $\beta = \sqrt{\beta_T^2 + \beta_z^2} < 1$. We also parameterize $R$, i.e. the transverse radius of fireball as $R = r_0\, exp\left(-\frac{z^2}{\sigma^2}\right)$ where $\sigma$ fixes the width of the matter distribution in the transverse direction [7, 9] and $z$ is the longitudinal coordinate of hadronic fluid element.

In our analysis, the contributions of various heavier hadronic resonances [10, 12] which decay after the thermal freeze-out of the hadronic matter has occurred are also taken into account. The invariant spectrum of a *given decay* product of a *given parent* hadron in the *local rest frame of a hadronic fluid element* is written as [7, 10, 12]:

$$E' \frac{d^3 N^{decay}}{d^3 p'} = \frac{1}{2p'} \left\{\frac{m_h}{p^*}\right\} \int_{E_-}^{E_+} dE_h \, E_h \left\{\frac{d^3 N_h}{d^3 p_h}\right\} \qquad (3)$$

where the subscript $h$ stand for the decaying (parent) hadron. The two body decay kinematics gives the **product** hadron's momentum and energy in the "rest frame of the decaying hadron" as $p^* = (E^{*2} - m^2)^{1/2}$ and $E^* = \frac{m_h^2 - m_j^2 + m^2}{2m_h}$ where $m_j$ indicates the mass of the other decay hadron produced along with the first one. The limits of integration are $E_\pm = \left\{\frac{m_h}{m^2}\right\} \{E'E^* \pm p'p^*\}$. The $E'(E_h)$ and $p'(p_h)$ are, respectively, the product (decaying parent) hadron's energy and momentum in the local rest frame of the hadronic fluid element.

## 3. Results and Discussions

We find, in our analysis, that the model calculations results fit the experimental data quite well. The experimental data are taken from the ALICE Collaboration for Pb + Pb collisions at $\sqrt{s_{NN}}$ = 2.76 TeV [13-15]. .

Over a fairly large $p_T$ range the hydrodynamical calculations show an approximate exponential behavior, whereas the tails of measured spectra show a significant deviation in the slope beyond 5 GeV at LHC. At RHIC this transition from exponential behavior takes place at $p_T \gtrsim 3$ GeV. The fraction of hadrons with very large $p_T$ ($\geq 3$ GeV at RHIC and $\geq 5$ GeV at LHC) is however small. We have considered the (maximum) $p_T$ range up to 5GeV in the present analysis. It is because that the statistical hydrodynamic calculations cannot describe the hadron spectra at such large transverse momenta. The hadrons detected in this region are essentially formed by the partons which are result of the hard processes. These originate from the direct *fragmentation* of high-energy partons of the colliding beams and therefore are not able to thermalize through the process of multiple collisions [16]. We therefore turn to softer hadrons which are assumed to be reasonably thermalized and form the bulk of the secondary matter produced.

The applicability region of hydrodynamics at LHC is therefore predicted to be for $p_T$ ≤ 4-5 GeV depending on the particle's mass. This range is wider than at RHIC [17]. The transverse momentum distributions are found to be sensitive to the values of the thermal/kinetic freeze-out temperature T and the transverse flow parameter $\beta_T^0$ whereas it is found to be *insensitive* to the change in the values of $\sigma$ in our model. In our analysis we have therefore fixed the value of the parameter $\sigma$ = 5.0. In our earlier analysis [2] of the RHIC data the value of $\sigma$ turned out to be nearly 4.2. However, it is expected to be large at the LHC energy. We have taken the values of *a* and *b* both to be zero for all the hadrons under the assumption of a baryon symmetric matter expected to be formed under the condition of a high degree of nuclear transparency in the nucleus-nucleus collisions at LHC energy. Unlike the previous works we have in our present analysis treated the index parameter *n* as a free parameter. The values of the parameters T, $\beta_T^0$ and *n* at freeze-out are determined by obtaining a best fit to a given hadron's transverse momentum spectrum.

In Figure 1, we have shown the transverse momentum spectra of protons and antiprotons. The values of the thermal freeze-out temperature T, the transverse flow parameter $\beta_T^0$ and the index parameter *n* for protons as well as antiprotons are found to be same, i.e. 102 MeV, 0.88 and 1.40, respectively, with a minimum $x^2/DoF$ of 0.61 for protons and 0.55 for anti-protons. The same values of the freeze-out parameters for protons and antiprotons indicate a simultaneous freeze-out of these particles in the dense hadronic medium.

The transverse momentum spectra for K$^+$ and K$^-$ shown in Figure 2 gives the value of (T, $\beta_T^0$, *n*) as (103 MeV, 0.89, 1.80) for Kaons and (105 MeV, 0.88, 1.80) for antiKaons. The minimum $x^2/DoF$ for both the two cases turns out to be 0.34. The almost similar freeze-out parameters obtained for protons, antiprotons, Kaons and antiKaons indicate a *near* simultaneous freeze-out of these particles.

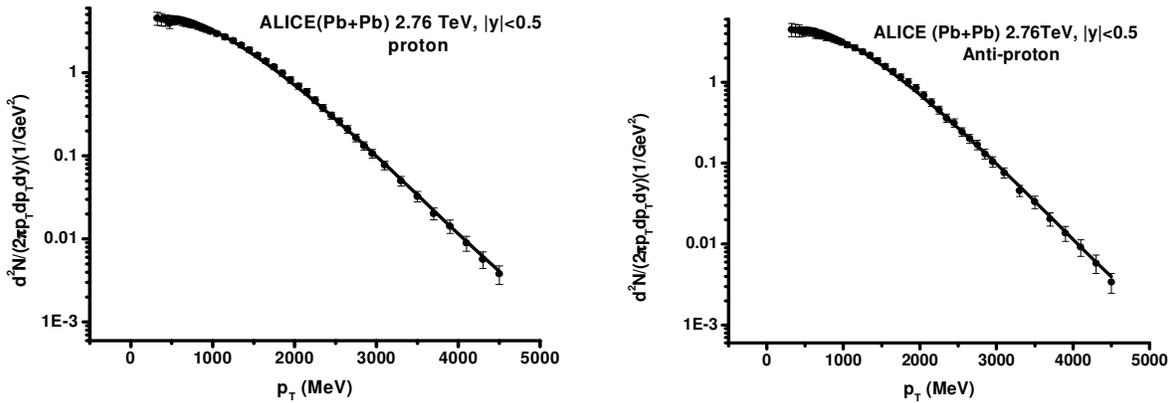

**Figure 1: Transverse momentum spectra of protons (left panel) and antiprotons (right panel) for the centrality class (0-5)%**

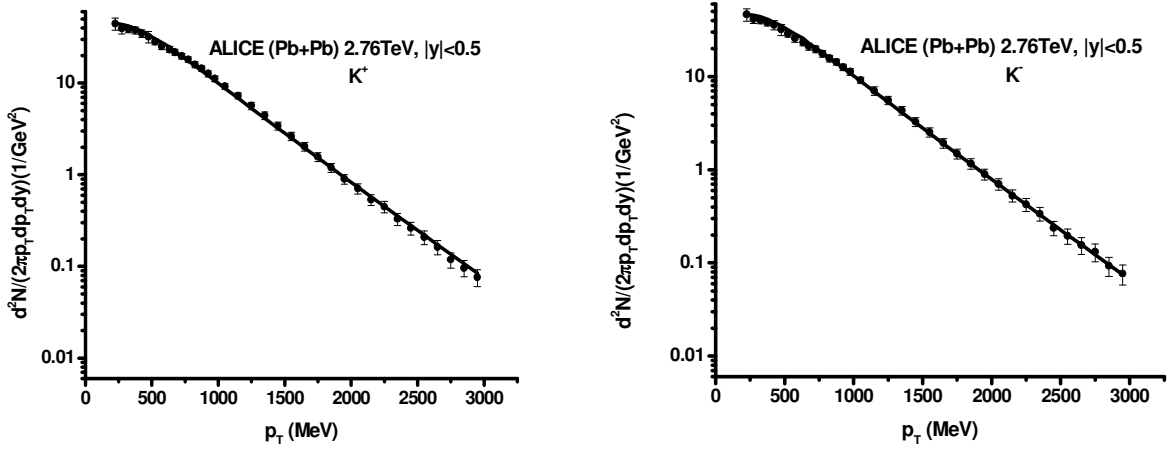

**Figure 2:** Transverse momentum spectra of K$^+$ (left panel) and K$^-$ (right panel) for the centrality class (0-5)%

The transverse momentum spectrum of neutral Kaon i.e. K$_s^0$ is shown in Figure 3. The values of T, $\beta_T^0$ and $n$ obtained from the spectra of K$_s^0$ are respectively 125 MeV, 0.84 and 1.61 with the minimum $x^2/DoF = 1.70$. The K$_s^0$ shows a larger thermal freeze-out temperature than the charged Kaons indicating its earlier freeze-out than K$^\pm$.

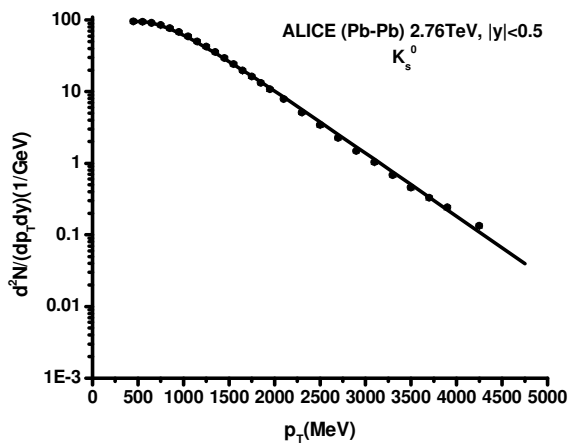

**Figure 3:** Transverse momentum spectra of K$_s^0$ for the centrality class (0-5)%.

The transverse momentum spectra of hyperons (i.e. $\Lambda$, $\Xi^-$, $\overline{\Xi^-}$, $\Omega$ and $\overline{\Omega}$) are shown in Figures 4, 5 and 6. The spectrum of $\Lambda$ gives the values of T, $\beta_T^0$ and $n$ as 127 MeV, 0.84 and 1.06, respectively, with a minimum $x^2/DoF = 0.52$. These values for $\Xi^-$ are found to be 133 MeV, 0.81 and 0.90 while for $\overline{\Xi^-}$ these are 149 MeV, 0.80 and 1.25, respectively. The parameters for $\Omega$ and $\overline{\Omega}$ are (155 MeV, 0.77 and 1.22) and (154 MeV, 0.77 and 1.23). The minimum $x^2/DoF$ for $\Xi^-$ and $\overline{\Xi^-}$ are 0.38 and 0.50 whereas for $\Omega$ and $\overline{\Omega}$ the minimum $x^2/DoF$ are 0.10 and 0.20, respectively. The relatively smaller values of minimum $x^2/DoF$ for $\Omega$s are due to larger experimental error bars.

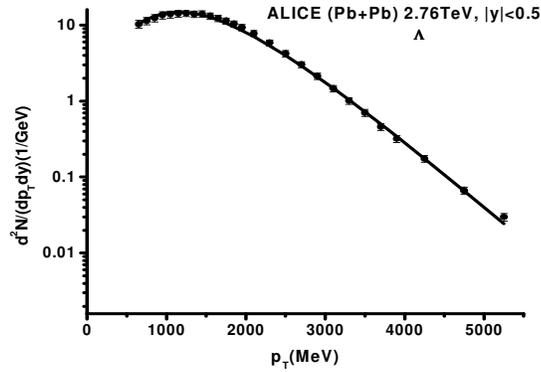

**Figure 4: Transverse momentum spectra of lambda $\Lambda$ for the centrality class (0-5)%.**

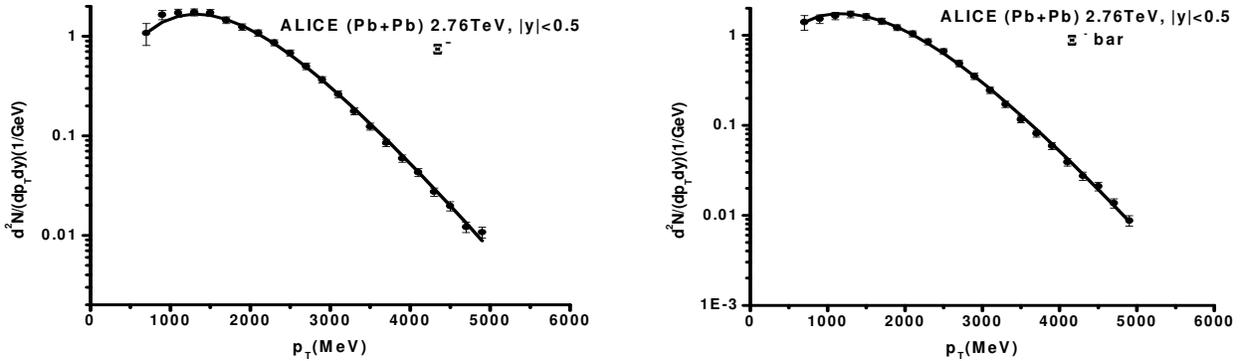

**Figure 5: Transverse momentum spectra of $\Xi^-$ (left panel) and $\overline{\Xi^-}$ (right panel) for the centrality class (0-10)%.**

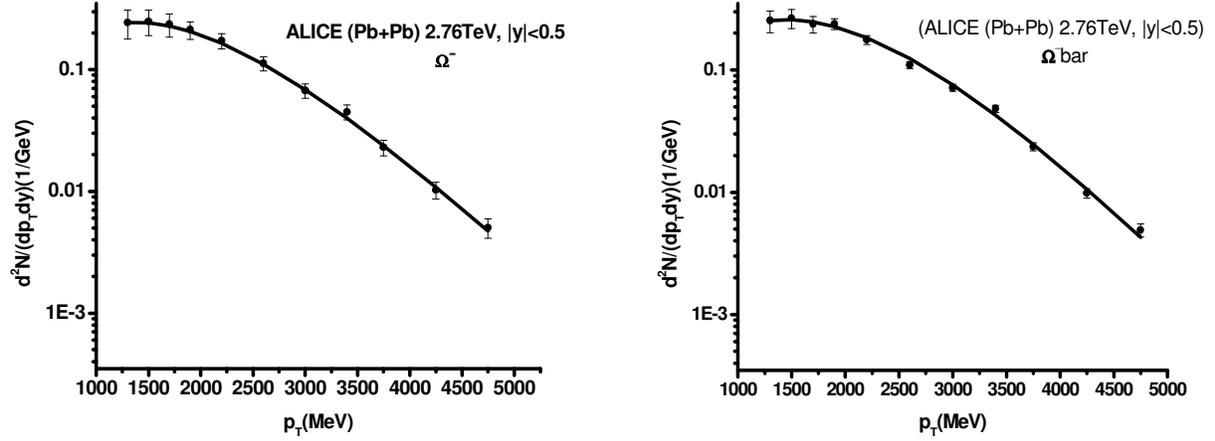

**Figure 6:** Transverse momentum spectra of $\Omega$ (left panel) and $\overline{\Omega}$ (right panel) for the centrality class (0-10)%.

The values of the thermal/kinetic freeze-out temperature T, the transverse flow parameter $\beta_T^0$ and the index parameter *n* for all the hadrons studied in this paper are again presented below in Table 1 to facilitate a proper comparison.

| Particle | T (MeV) | $\beta_T^0$ | n | $x^2/DoF$ |
|---|---|---|---|---|
| $p$ | 102 | 0.88 | 1.40 | 0.61 |
| $\bar{p}$ | 102 | 0.88 | 1.40 | 0.55 |
| $K^+$ | 103 | 0.89 | 1.80 | 0.34 |
| $K^-$ | 105 | 0.88 | 1.80 | 0.34 |
| $K_S^0$ | 125 | 0.84 | 1.61 | 1.70 |
| $\Lambda$ | 127 | 0.84 | 1.06 | 0.52 |
| $\Xi^-$ | 133 | 0.81 | 0.90 | 0.38 |
| $\overline{\Xi}$ | 149 | 0.80 | 1.25 | 0.48 |
| $\Omega$ | 155 | 0.77 | 1.22 | 0.10 |
| $\overline{\Omega}$ | 154 | 0.77 | 1.23 | 0.20 |

**Table 1 :** Freeze-out parameters of various hadrons obtained from their Transverse Momentum spectra.

It is evident from Table 1 that the lighter particles, that is, (anti)protons and Kaons, exhibit a lower thermal freeze-out temperature and a higher surface transverse expansion velocity compared to the heavy (multi)strange hyperons. The reason for this can be attributed to an early freeze-out for the massive particles (hyperons) when the thermal temperature is high and the collective flow is in the early stage of development and consequently $\beta_T^0$ is small. The early freeze-out of these particles is due to their smaller cross-section with the hadronic matter.

A comparison with a similar fit to the RHIC data [2, 6] shows that for the most central collisions the flow velocity increases significantly at LHC, reaching almost 0.9 and that the kinetic freeze-out temperature drops below the one at RHIC. For RHIC [2] the values of T and $\beta_T^0$ were found to be in the range 163– 188 MeV and 0.58 – 0.67, respectively.

## 4. Conclusion

The transverse momentum spectra of the hadrons p, $\bar{p}$, $K^+$, $K^-$, $K_S^0$, $\Lambda$, $\Omega$, $\bar{\Omega}$, $\Xi^-$ and $\bar{\Xi}$ are fitted quite well by using our model. The assumption of vanishing chemical potential at midrapidity shows the effects of almost complete transparency in Pb+Pb collisions at LHC energy of 2.76 TeV. We also observe an earlier freeze-out of hyperons as compared to lighter mass particles i.e. Kaons and protons. The protons, antiprotons, Kaons and anti Kaons have similar freeze-out conditions, which indicate their near simultaneous freeze-out from the dense hadronic medium. The larger values of $\beta_T^0$ at the LHC energy as compared to those at RHIC, indicates a stronger flow effect present in the system at LHC.

## Acknowledgements

Inam-ul Bashir is thankful to University Grants Commission for awarding the Basic Scientific Research (BSR) Fellowship. Riyaz Ahmed Bhat is grateful to Council of Scientific and Industrial Research, New Delhi for awarding Senior Research Fellowship. Saeed Uddin is grateful to University Grants Commission for financial assistance.